\documentclass[amsmath, amssymb,10pt,aps,prb,twocolumn,notitlepage,showpacs,superscriptaddress]{revtex4-1}
\usepackage{graphicx}
\usepackage{calc}
\usepackage{braket}
\usepackage{ulem}   
\usepackage{slashed}
\usepackage{wasysym}
\usepackage{siunitx}
\usepackage{dsfont}
\usepackage{amsthm,amsmath,amsfonts,amssymb,verbatim,color}
\usepackage{graphicx}
\usepackage{subfigure}
\usepackage{bm}
\usepackage{epsfig,slashed}
\usepackage[T1]{fontenc}
\usepackage[colorlinks=true,citecolor=blue,linkcolor=blue,urlcolor=blue]{hyperref}

\newcommand{\icm}{\ensuremath{~\textrm{cm}^{-1}}}

\begin{document}

\title{Linear and nonlinear optical responses in the chiral multifold semimetal RhSi}

\author{Zhuoliang Ni}
\affiliation{Department of Physics and Astronomy, University of Pennsylvania, Philadelphia, Pennsylvania 19104, USA}
\author{B. Xu}
\affiliation{University of Fribourg, Department of Physics and Fribourg Center for Nanomaterials, Chemin du Mus\'{e}e 3, CH-1700 Fribourg, Switzerland}
\author{M.-\'{A}. S\'{a}nchez-Mart\'{i}nez}
\affiliation{Univ. Grenoble Alpes, CNRS, Grenoble INP, Institut N\'eel, 38000 Grenoble, France}
\author{Y. Zhang}
\affiliation{Max-Planck-Institut fur Chemische Physik fester Stoffe, 01187 Dresden, Germany}
\affiliation{Department of Physics, Massachusetts Institute of Technology, Cambridge, Massachusetts 02139, USA}
\author{K. Manna}
\affiliation{Max-Planck-Institut fur Chemische Physik fester Stoffe, 01187 Dresden, Germany}
\affiliation{Department of Physics, Indian Institute of Technology Delhi, Hauz Khas, New Delhi-110016, India}
\author{C. Bernhard}
\affiliation{University of Fribourg, Department of Physics and Fribourg Center for Nanomaterials,
Chemin du Mus\'{e}e 3, CH-1700 Fribourg, Switzerland}
\author{J. W. F. Venderbos}
\affiliation{Department of Physics and Astronomy, University of Pennsylvania, Philadelphia, Pennsylvania 19104, USA}
\affiliation{Department of Physics \& Department of Materials Science and Engineering, Drexel University, Philadelphia, Pennsylvania 19104, USA}
\author{F. de Juan}
\affiliation{Donostia International Physics Center, P. Manuel de Lardizabal 4, 20018 Donostia-San Sebastian, Spain}
\affiliation{IKERBASQUE, Basque Foundation for Science, Maria Diaz de Haro 3, 48013 Bilbao, Spain}
\author{C. Felser}
\affiliation{Max-Planck-Institut fur Chemische Physik fester Stoffe, 01187 Dresden, Germany}
\author{A. G. Grushin}
\email{adolfo.grushin@neel.cnrs.fr}
\affiliation{Univ. Grenoble Alpes, CNRS, Grenoble INP, Institut N\'eel, 38000 Grenoble, France}
\author{Liang Wu}
\email{liangwu@sas.upenn.edu}
\affiliation{Department of Physics and Astronomy, University of Pennsylvania, Philadelphia, Pennsylvania 19104, USA}

\date{\today}

\begin{abstract}
Chiral topological semimetals are materials that break both inversion and mirror symmetries. They host interesting phenomena such as the quantized circular photogalvanic effect (CPGE) and the chiral magnetic effect.   In this work, we report a comprehensive theoretical and experimental analysis of the linear and non-linear optical responses of the chiral topological semimetal RhSi, which is known to host multifold fermions. We show that the characteristic features of the optical conductivity, which display two distinct quasi-linear regimes above and below 0.4 eV, can be linked to excitations of different kinds of multifold fermions. The characteristic features of the CPGE, which displays a sign change at 0.4 eV and a large non-quantized response peak of around 160 $\mu \textrm{A V}^{-2}$ at 0.7 eV, are explained by assuming that the chemical potential crosses a flat hole band at the Brillouin zone center.  Our theory predicts that, in order to observe a quantized CPGE in RhSi, it is necessary to increase the chemical potential as well as the quasiparticle lifetime. More broadly our methodology, especially the development of the broadband terahertz emission spectroscopy, could be widely applied to study photo-galvanic effects in noncentrosymmetric materials and in topological insulators in a contact-less way and accelerate the technological development of efficient infrared detectors based on topological semimetals. 

\end{abstract}

\pacs{}
\maketitle

\textbf{Introduction} 

The robust and intrinsic electronic properties of topological metals---a class of quantum materials---can potentially protect or enhance useful electromagnetic responses\cite{MurakamiNJP07,WanPRB2011,BurkovPRL2011,armitageRMP2018}. However, direct and unambiguous detection of these properties is often challenging. For example, in Dirac semimetals such as Cd$_3$As$_2$ and Na$_3$Bi\cite{wangPRB2012, wangPRB2013} two doubly degenerate bands cross linearly at a single point, the Dirac point, and this crossing is protected by rotational symmetry \cite{liuScience2014, neupaneNatComm2014, liuNatMat2014}. A Dirac point can be understood as two coincident topological crossings with equal but opposite topological charge\cite{armitageRMP2018}, and as a result, the topological contributions to the response to external probes cancel in this class of materials, rendering external probes insensitive to the topological charge.

 Weyl semimetals may offer an alternative, as this class of topological metals is defined by the presence of isolated twofold topological band crossings, separated in momentum space from a partner crossing with opposite topological charge. This requires the breaking of either time-reversal or inversion symmetry. The Weyl semimetal phases discovered in materials of the transition monopnictide family such as TaAs\cite{WengPRX2015, HuangNatComm2015, XuScience2015, LvPRX2015, lvNatPhys2015,xuNatphys2015, YangNatPhys2015,xuNatComm2016} lack inversion symmetry, which allows for nonzero second-order nonlinear optical responses and has motivated the search for topological responses using techniques of nonlinear optics. This search has resulted in the observation of giant second harmonic generation\cite{wuNatPhys2017,patankarPRB2018}, as well as interesting photo-galvanic effects\cite{maNatPhys2017, sunCPL2017, osterhoudtNatMat2019, siricaPRL2019, gaoNatComm2020}. However, neither response can be directly attributed to the topological charge of a single band crossing, since mirror symmetry---present in most known Weyl semimetals---imposes that charges with opposite sign lie at the same energy and thus contribute equally~\cite{patankarPRB2018, zhangPRB2018}. This is similar for other types of Weyl semimetal materials, such as type-II Weyl semimetals\cite{soluyanovNature2015}. Type-II Weyl semimetals display open Fermi surfaces to lowest order in momentum\cite{soluyanovNature2015, wangPRL2016, changNatComm2016, huangNatComm2016, belopolskiNatComm2016, dengNatPhys2016, jiangNatComm2017}, giving rise to remarkable photogalvanic effects\cite{yangarXiv2017, limPRB2018, jiNatMat2019, maNatMat2019, wangNatComm2019}, but not directly linked to
their topological charge.

 Materials with even lower symmetry can hold the key to measuring the topological charge directly. Chiral topological metals do not possess any inversion or mirror symmetries\cite{Manes:2012fi,wiederPRL2016,bradlynScience2016,changNatMat2018}, and as a result, the topological band crossings do not only occur at different momenta but also at different energies, making them accessible to external probes. Notably, the circular photogalvanic effect (CPGE), i.e., the part of the photocurrent that reverses sign with the sense of polarization, was predicted to be quantized in chiral Weyl semimetals\cite{dejuanNatComm2017}. However, chiral Weyl semimetals with sizable Weyl node separations, such as SrSi$_2$\cite{Huang:2016is}, have not been synthesized as single crystals.

Recently, a class of chiral single crystals has emerged as a promising venue for studying topological semimetallic behavior deriving from topological band crossings. Following a theoretical prediction\cite{Manes:2012fi,wiederPRL2016,bradlynScience2016,changPRL2017, tangPRL2017, changNatMat2018}, experimental evidence provided proof that a family of silicides, including CoSi \cite{sanchez2019topological, raoNature2019, takanePRL2019} and RhSi \cite{sanchez2019topological}, hosts topological band crossings with nonzero topological charge at which more than two bands meet. Such band crossings, known as multifold nodes, may be viewed as generalizations of Weyl points and are enforced by crystal symmetries. These materials are good candidates to study signatures of topological excitations in optical conductivity measurements, as the Lifshitz energy which separates the topological from the trivial excitations is on the order of $\sim 1$ eV. In contrast, the Lifshitz energy in previous Dirac/Weyl semimetals such as Cd$_3$As$_2$ \cite{wangPRB2013}, Na$_3$Bi \cite{wangPRB2012} and TaAs \cite{WengPRX2015,HuangNatComm2015} is below 100 meV.

The prediction of a quantized CPGE was extended to materials in this class, specifically to RhSi in space group 198\cite{changPRL2017,flickerPRB2018,deJuan:2020jm}. These materials display a protected three-band crossing of topological charge $2$, known as a threefold fermion at the zone center and a protected double Weyl node of opposite topological charge at the zone boundary. In RhSi, theory predicts that below 0.7 eV, only the $\Gamma$ point is excited, resulting in a CPGE plateau when the chemical potential is above the threefold node\cite{changPRL2017,flickerPRB2018,deJuan:2020jm}. Above 0.7 eV, the R point contribution of opposite charge compensates it, resulting in a vanishing CPGE at large frequencies\cite{changPRL2017,flickerPRB2018,deJuan:2020jm}. The predicted energy dependence above 0.5 eV  is qualitatively consistent with a recent experiment in RhSi performed within photon energies ranging between 0.5 eV to 1.1 eV \cite{reesarxiv2019}.

Despite this preliminary progress, the challenge to determine if and how quantization can be observed in practice in these materials has remained unanswered, largely since multiple effects such as the quadratic correction \cite{flickerPRB2018,deJuan:2020jm} and short hot-carrier lifetime \cite{dejuanNatComm2017,Konig2017} can conspire to destroy it. Furthermore, thus far, the experimental signatures of the existence of multifold fermions have been limited to band structure measurements \cite{sanchez2019topological, raoNature2019, takanePRL2019}. A comprehensive understanding of the linear \cite{maulanaPRR2020} and nonlinear optical responses \cite{reesarxiv2019}, targeting the energy range where the multifold fermions dominate optical transitions and transport signatures, is still lacking.

In this work we report the measurement of the linear and non-linear response of RhSi, analyzed using different theoretical models of increasing complexity, and provide a consistent picture of {\it (i)} the way in which multifold fermions manifest in optical responses, and {\it (ii)} how quantization can be observed. We performed optical conductivity measurements from 4 meV to 6 eV and 10 K to 300 K, as well as terahertz (THz) emission spectroscopy with incident photon energy from 0.2 eV to 1.1 eV at 300 K. Our optical conductivity measurements, combined with tight-binding and \textit{ab-initio} calculations, show that interband transitions  $\lesssim0.4$ eV are mainly dominated by the vertical transitions near the multifold nodes at the Brillouin zone center, the $\Gamma$ point. We found that the transport lifetime is relatively short in RhSi, $\leq 13$ fs at 300 K and $\leq 23$ fs at 10 K. The measured CPGE response shows a sign change and no clear plateau. Our optical conductivity and CPGE experiments are reasonably well reproduced by tight-binding and first-principle calculations when the chemical potential lies below the threefold node at the $\Gamma$ point, crossing a relatively flat band, and when the hot-carrier lifetime is chosen to be $\approx 4$-$7$ fs.  We argue that these observations are behind the absence of quantization. Our \textit{ab-initio} calculation predicts that a quantized CPGE could be observed by increasing the electronic doping by $100$ meV with respect to the chemical potential in the current generation of samples~\cite{sanchez2019topological,reesarxiv2019,maulanaPRR2020}, if it is accompanied by an improvement in the sample quality that can significantly increase the hot carrier lifetime.\\ 

\textbf{Results and Discussion}

\begin{figure*}
\includegraphics[width=0.9\textwidth]{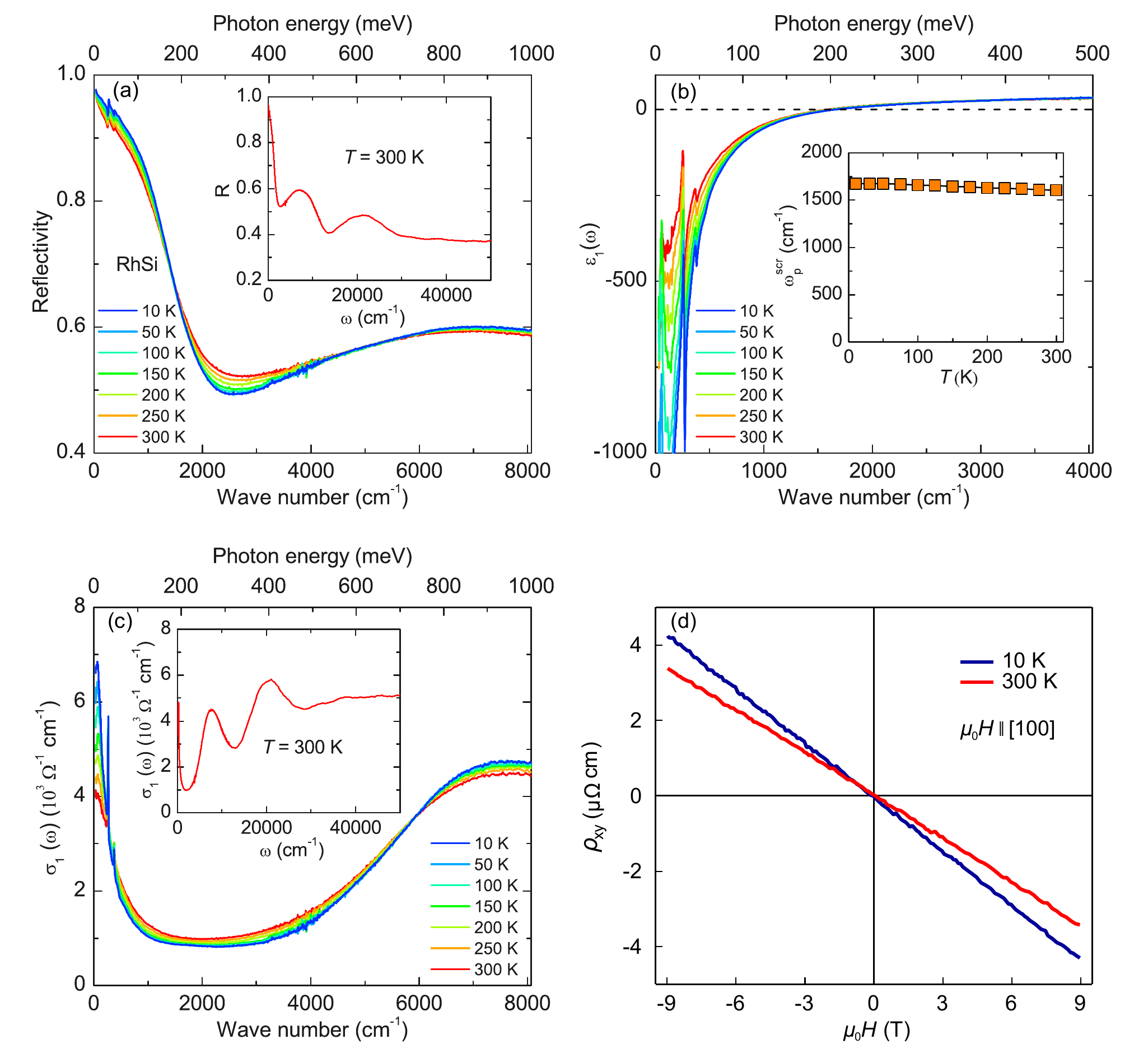}
\caption{ (color online) \textbf{Optical conductivity and Hall resistivity of RhSi.} (a) Temperature dependence of the reflectivity spectra up to 8\,000\icm\ of RhSi. Inset shows the room temperature spectrum up to 50\,000\icm. (b) Temperature dependence of the real part of the dielectric function $\varepsilon_1(\omega)$. Inset shows screened plasma frequency of free carriers obtained from the zero crossings of $\varepsilon_1(\omega)$. (c) Optical conductivity of RhSi up to 8\,000\icm\ at different temperatures. Inset: Optical conductivity shown up to 50\,000\icm\ at room temperature. (d) Hall resistivity of RhSi at 10 K and 300 K.}
\label{Fig1}
\end{figure*}

\textbf{Optical conductivity measurement} 

The measured frequency-dependent reflectivity $R(\omega)$ by a FTIR spectrometer (see Methods) is shown in Fig.~\ref{Fig1}(a) in the frequency range from 0 to 8\,000\icm\ for several selected temperatures. (1 meV corresponds to 8.06\icm and 0.24 THz.) $R(\omega)$ at room temperature is shown over a much larger range up to 50\,000\icm\ in the inset. In the low-frequency range, $R(\omega)$ is rather high and has a $R = 1 - A\sqrt{\omega}$ response characteristic of a metal in the Hagen-Rubens regime. Around 2\,000\icm\ a temperature-independent plasma frequency is observed in the reflectivity. For $\omega >$ 8000\icm\ the reflectivity is approximately temperature independent.

The results of the Kramers-Kronig analysis of $R(\omega)$ are shown in Figs.~\ref{Fig1}(b) and \ref{Fig1}(c) in terms of the real part of the dielectric function $\varepsilon_1(\omega)$ and the real part of optical conductivity $\sigma_1(\omega)$. At low frequencies, $\varepsilon_1(\omega)$ is negative, a defining property of a metal. With increasing photon energy $\omega$, $\varepsilon_1(\omega)$ crosses zero around 1\,600\icm\ and reaches values up to 33  around 4\,000\icm. The crossing point, where $\varepsilon_1(\omega)=0$, is related to the screened plasma frequency $\omega_p^{scr}$ of free carriers. As shown by the inset of Fig.~\ref{Fig1}(b), $\omega_p^{scr}$ is almost temperature independent. Similar temperature dependence and values of $\omega_p^{scr}$ have been recently reported in another work for RhSi~\cite{maulanaPRR2020}, indicating similar large carrier densities and small transport lifetime in RhSi samples.

Fig.~\ref{Fig1}(c) shows the temperature dependence of $\sigma_1(\omega)$ for RhSi. Overall, ${\sigma}_1(\omega)$ is dominated by a narrow Dude-like peak in the far-infrared region, followed by a relativety flat tail in the frequency region between  1\,000 and 3\,500\icm. As the temperature decreases, the Drude-like peak narrows with a concomitant increase of the low-frequency optical conductivity. In addition, the inset shows the $\sigma_1(\omega)$ spectrum at room temperature over the entire measurement range, in which the high-frequency $\sigma_1(\omega)$ is dominated by two interband transition peaks around 8\,000\icm and 20\,000\icm.

\begin{figure*}
\includegraphics[width=\textwidth]{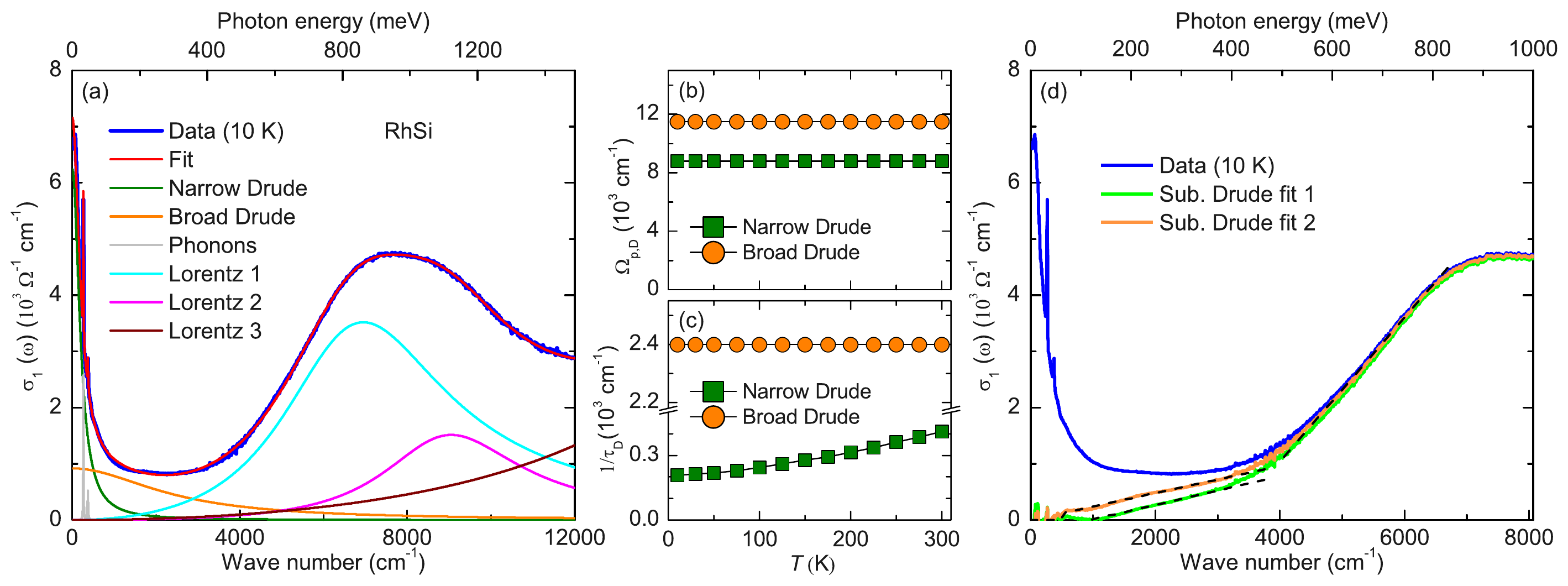}
\caption{(color online) \textbf{Drude and interband optical responses in RhSi.} (a) Optical conductivity spectrum of RhSi up to 12\,000\icm\ (1.5~eV) at 10~K. The thin red line through the data is the Drude-Lorentz fitting result, which consists of the contributions from a narrow Drude peak (green line), a broad Drude peak (orange line) and several Lorentz terms that accounts for the phonons (gray) and the interband transitions (light blue, magenta, and wine lines). Temperature dependence of (b) the plasma frequency $\Omega_{p,D}$ and (c) the transport scattering rate $1/\tau_D$ of the Drude terms. (d) Optical conductivity spectrum of RhSi at 10 K, and the corresponding spectrum after the Drude response and the sharp phonon modes have been subtracted. Black dashed lines are eye guidance for different quasi-linear regimes.}
\label{fig2}
\end{figure*}

To perform a quantitative analysis of the optical data at low frequencies, we fit the ${\sigma}_1(\omega)$ spectra with a Drude-Lorentz model
\begin{equation}
\label{DrudeLorentz}
\sigma_{1}(\omega)=\frac{2\pi}{Z_{0}}\biggl[\sum_{j}\frac{\Omega^{2}_{pD,j}}{\omega^{2}\tau_{D,j} + \frac{1}{\tau_{D,j}}} + \sum_{k}\frac{\gamma_{k}\omega^{2}S^{2}_{k}}{(\omega^{2}_{0,k} - \omega^{2})^{2} + \gamma^{2}_{k}\omega^{2}}\biggr],
\end{equation}
where $Z_{0}$ is the vacuum impedance. The first sum of Drude-terms describes the response of the itinerant carriers in the different bands that are crossing the Fermi-level, each characterized by a plasma frequency $\Omega_{pD,j}$ and a transport scattering rate (Drude peak width) $1/\tau_{D,j}$. The second term contains a sum of Lorentz oscillators, each with a different resonance frequency $\omega_{0,k}$, a line width $\gamma_k$, and an oscillator strength $S_k$. The corresponding fit to the conductivity at 10~K (thick blue line) using the function of Eq.~\eqref{DrudeLorentz} (red line) is shown in Fig.~\ref{fig2}(a) up to 12\,000\icm. As shown by the thin colored lines, the fitting curve is composed of two Drude terms with small and large transport scattering rates, respectively, and several Lorentz terms that account for the phonons at low energy and the interband transitions at higher energy. Fits of the $\sigma_1(\omega)$ curves at all the measured temperatures return the temperature dependence of the fitting parameters. Figure~\ref{fig2}(b) shows the temperature dependence of the plasma frequencies $\Omega_{p,D}$ of the two Drude terms, which remain constant within the error bar of the measurement, indicating that the band structure hardly changes with temperature. Figure~\ref{fig2}(c) displays the temperature dependence of the corresponding transport scattering rates $1/\tau_D$ of the two Drude terms. The transport scattering rate of the broad Drude term remains temperature independent, while that of the narrow Drude decreases at low temperature. Note that the temperature dependence of the Drude responses appears to be slightly stronger than in Ref. \onlinecite{maulanaPRR2020} probably due to a slightly better crystal quality in our studies.

\begin{figure*}
\includegraphics[width=\textwidth]{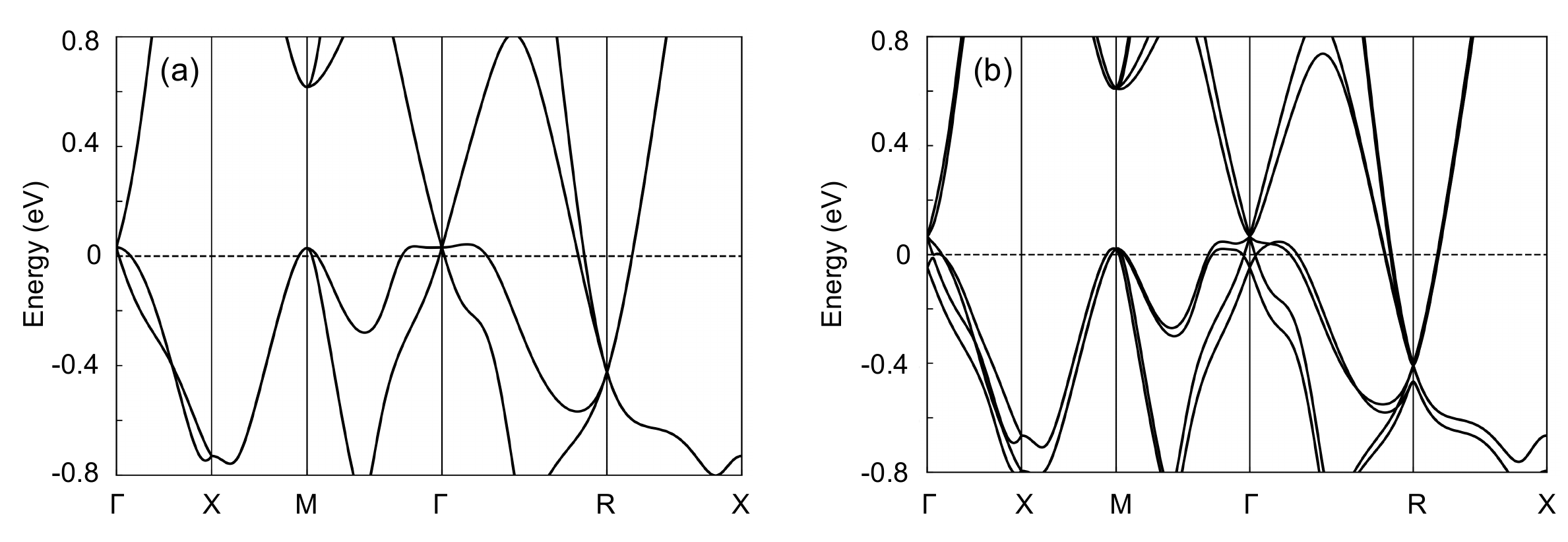}
\caption{\label{fig3}  \textbf{DFT band structure of RhSi.}
(a) without spin orbit coupling, and (b) with spin orbit coupling. It hosts a threefold fermion at $\Gamma$ and a fourfold double Weyl fermion at $R$ without spin orbit coupling. They split into a spin-3/2 node and a Weyl node at $\Gamma$, and a sixfold double spin-1 node and a twofold Kramers node at $R$, when spin orbit coupling is included.
\label{fig3}}
\end{figure*}

The need for two Drude terms indicates that RhSi has two types of charge carriers with very different transport scattering rates. Such a two-Drude fit is often used to describe the optical response of multiband systems. Prominent examples of such multiband materials are the iron-based superconductors ~\cite{Wu2010PRB,Dai2013PRL,Xu2019PRB}. As we discuss below, in the case of RhSi two main pockets are expected to cross the Fermi level, centered around the $\Gamma$ (heavy hole pocket) and the $R$ point (electron pocket)~\cite{changPRL2017,tangPRL2017}. (See the band structure in Fig. \ref{fig3}.) Note that there might be a small hole pocket at the $M$ point as well. Accordingly, the two-Drude fit can most likely be assigned to the intraband response around the $\Gamma$ (broad Drude term) and $R$ (narrow Drude term) points of the Brillouin zone because the effective mass of the holes from the flat bands at $\Gamma$ is much heavier than the electrons at $R$. This is further supported by the observation of dominant electron contribution in Hall resistivity measurement on a typical RhSi sample, which is linear as a function of magnetic field up to 9 T, as shown in Fig. \ref{Fig1}(d). A third Drude peak for the pocket at $M$ could be included but its contribution must be very small as the two pockets at $\Gamma$ and $R$ are much larger. Note that the two Drude terms could also come from two scattering processes with different scattering rates \cite{maulanaPRR2020}. We use the transport lifetime of the narrow Drude peak as the upper bound and estimate that  transport lifetime is $\leq 13$ fs at 300 K and $\leq 23$ fs at 10 K, consistent with previous studies \cite{reesarxiv2019, maulanaPRR2020}.

Having examined the evolution of the two-Drude response with temperature, we next investigate the $\sigma_1(\omega)$ spectrum associated with interband transitions. To single out this contribution, we show in Fig.~\ref{fig2}(d) the $\sigma_1(\omega)$ spectra, after subtracting the two-Drude response and the sharp phonon modes. With the subtraction of two Drude peaks with transport scattering rates of 200\icm\ and 2\,400\icm\ (Drude fit 1 in Fig.~\ref{fig2}(d)), we reveal a quasi-linear behavior of $\sigma_1(\omega)$ in the low-frequency regime (up to about 3\,500\icm). Such behavior is a strong indication for the presence of three-dimensional linearly dispersing bands near the Fermi level~\cite{SanchezMartinez:2019he}. Indeed, from band structure calculations (see Fig.~\ref{fig3}), we see that this low-energy quasi-linear interband conductivity ($\omega <$ 3\,500\icm) could be attributed to the interband transitions around the $\Gamma$ point. At higher energy, the interband contributions around the $R$ point become allowed and can be responsible for the second quasi-linear interband conductivity region (3\,500\icm\ $< \omega <$ 6\,500\icm). At $\omega >$ 6\,500\icm, the optical conductivity flattens and forms a broad maximum around 8\,000\icm. From Fig. \ref{fig2}(a) we see that this maximum is a consequence of the Lorentzian peak around 0.85 eV (light blue) and around 1.1 eV (magenta). As analyzed by density-functional theory below, the peak around 0.85 eV is most likely attributed to broadened interband transitions centered at the $M$ point, which was previously systematically studied in CoSi~\cite{xuarXiv2020}. See more discussion in the calculation below. Note that our interpretation of the peak is different from Ref. \onlinecite{maulanaPRR2020}. 

\begin{figure*}
\includegraphics[width=\textwidth]{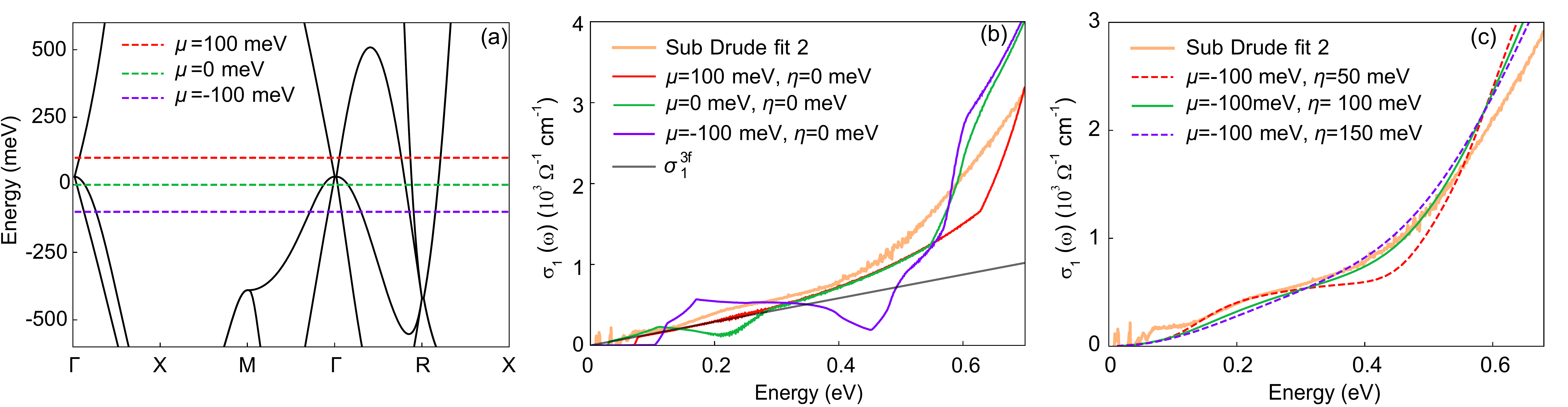}
\caption{\label{fig4} \textbf{Four-band model.} (a) Band structure obtained using a tight binding model for RhSi without spin-orbit coupling. (b) Optical conductivity of RhSi for different chemical potentials  $\mu$ without disorder broadening.  The $\sigma^{\mathrm{3f}}_1$ conductivity by low-energy linear model  is shown as a solid gray line. (c) Optical conductivity of RhSi calculated with different disorder related broadenings, $\eta$, for $\mu=-100$ meV (see (b)) using Eq.~\eqref{eqn:optcond}. The solid green curve has the same parameters ($\mu=-100$ meV, $\eta=100$ meV)  as the tight-binding CPGE calculation (orange curve in Fig.~\ref{fig6}(c)).}
\end{figure*}

\begin{figure*}
\includegraphics[width=\textwidth]{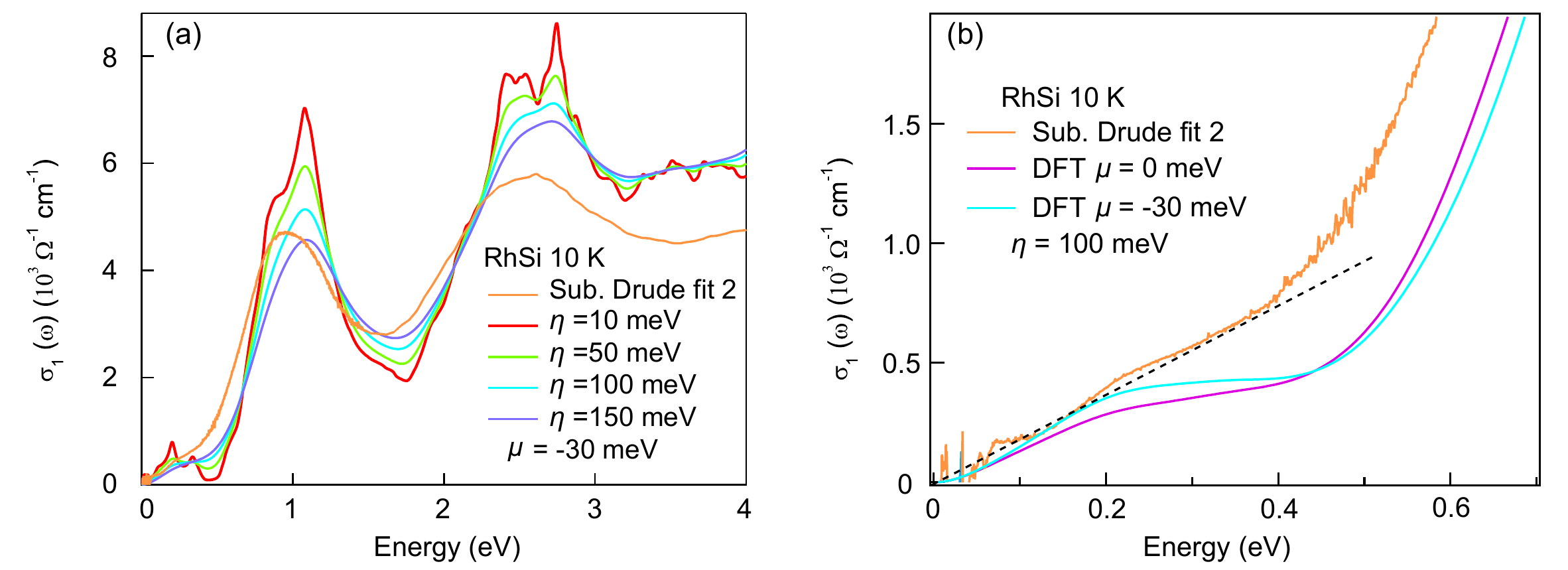}
\caption{\label{fig5}  
\textbf{Calculated optical conductivity by DFT.} (a) Full frequency-range optical conductivity comparing different broadening factors $\eta$, with fixed $\mu=-30$ meV. The $\eta=100$ meV curve shows a relatively good agreement up to 4 eV.
(b) Low-energy optical conductivity data compared to DFT results calculated at different chemical potentials $\mu$, and a disorder broadening $\eta=100$ meV. The black dashed line is a guide to the eye, highlighting the small deviations from linearity, and a small shoulder in the data around 200 meV.
\label{fig5}}
\end{figure*}

Before analyzing these further, it is important to note that the fit to the broader Drude peak might suffer from more uncertainty than that of the narrow Drude peak. Small changes in its width might result in appreciable changes when subtracting it from the full data set to obtain the interband response. Here we use a different method, which does not make use of Drude Lorentz fits since the low-frequency tails of Lorentian terms might also look quasi-linear. Instead, we fit and then subtract the two Drude and two phonon terms directly. By subtracting the broad Drude peak this time with a smaller transport scatter rate of 1\,350\icm\ (Drude fit 2 in Fig.~\ref{fig2}(d)), the onset frequency at which the interband conductivity emerges decreases and the magnitude of $\sigma_1$ below 4\,000\icm\  increases, with respect to the Drude fit 1. However, the resulting slope below 3\,500 \icm\ is not significantly modified as the wide Drude response contributes as a flat background in this regime. Note that another recent study used a similar method and also found that the quasi-linear behavour is robust within the uncertainty of the fit parameters \cite{maulanaPRR2020}. Therefore, we conclude that the low-frequency quasi-linear conductivity is contributed from interband excitations and will be analyzed next in our theoretical modeling. \\

\textbf{Optical conductivity calculation} 

To gain insight into the low-frequency interband optical conductivity of RhSi, we now put these predictions to the test based on a low-energy linearized model around the $\Gamma$ point, a four-band tight-binding model, and \textit{ab-initio} calculations (see Methods).

We start from simpler linear and tight-binding models to explain the two low-energy quasi-linear behavior.  The band structure of RhSi calculated using Density Functional Theory (DFT) is shown in Fig.~\ref{fig3}.  The band structure calculations of Fig.~\ref{fig3} suggest that at low frequencies, $\omega \lesssim 0.4$ eV, the optical conductivity is dominated by interband transitions close to the $\Gamma$ point as the $R$ point is 0.4 eV below the chemical potential. To linear order in momentum, the middle band at $\Gamma$ is flat (see Fig.~\ref{fig3}(a)), and the only parameter is the Fermi velocity of the upper and lower dispersing bands $v_\mathrm{F}$. The optical conductivity of a threefold fermion is $\sigma^{\mathrm{3f}}_1 = e^2\omega/ (12\pi \hbar v_\mathrm{F})$ ~\cite{SanchezMartinez:2019he}and is plotted in Fig. \ref{fig4}(b)(solid gray line).  We observe that the low energy $\sigma^{\mathrm{3f}}_1$ shows a smaller average slope compared to the low-energy part of our experimental data. In addition, a purely linear conductivity is insufficient to describe a small shoulder at around $200$ meV (see a more zoomed-in data in Fig.~\ref{fig5}(b) compared to the linear guide to the eye).

The position of the chemical potential in Fig.~\ref{fig3} indicates that the deviations from linearity of the central band at $\Gamma$ can play a significant role in optical transitions. To include them we use a four-band tight-binding lattice model that incorporates the lattice symmetries of space-group 198~\cite{changPRL2017,flickerPRB2018}.  The resulting tight-binding band structure is shown in Fig.~\ref{fig4}(a) by following the method developed in Ref. \onlinecite{flickerPRB2018}. (See Methods).  In Fig.~\ref{fig4}(b) we compare  the tight-binding model with different chemical potentials indicated in Fig.~\ref{fig4}(a) and the experimental optical conductivity in the interval $\hbar \omega \in [0,0.7]$ eV  (sub. Drude fit 2), as the extrapolation is expected to go through zero ~\cite{SanchezMartinez:2019he}. By choosing different chemical potentials below and above the node  without yet including a hot-carrier scattering time $\tau$, we observe that if the chemical potential is below the threefold node (see $\mu=0,-100$ meV curves in Fig.~\ref{fig4} (b)), a peak appears (around $200$ meV for $\mu=-100$ meV), followed by a dip in the optical conductivity at larger frequencies, before the activation of the transitions centered at the $R$ point.

The peak-dip feature observed in the optical conductivity can be traced back to the allowed optical transitions and the curvature of the middle band. When the Fermi level lies below the node, the interband transitions with the lowest activation frequency connect the lower to the middle threefold band at $\Gamma$. Increasing the frequency could activate transitions between the bottom and upper threefold bands, allowed by quadratic corrections, but these are largely suppressed due to the selection rules as the change of angular momentum between these bands is $2$\cite{SanchezMartinez:2019he}. Because of the curvature of the middle band, the transitions connecting the lower and the middle threefold bands die out as frequency is increased further, resulting in the peak-dip structure visible in Fig.~\ref{fig4} (b). Since the curvature of the middle band is absent by construction in the linear model but captured by the tight-binding model, it is only the latter model that shows a conductivity peak-dip.  As a side remark, we observe that the transitions involving the R point bands  activate at lower frequencies as the chemical potential is decreased.

Although placing the chemical potential below the $\Gamma$ node results in a marked peak around 0.2 eV, it is clearly sharper, and overshoots compared to the data, for which the sudden drop at frequencies above the peak is also absent. It is likely that this drop is masked by the finite and relatively larger disorder related broadening $\eta=\hbar/\tau$. This scale is expected to be large for RhSi given the broad nature of the low-energy Drude peak width in Eq.~\ref{eqn:optcond}. Note that $\tau$ is the hot-carrier lifetime, which is different from the transport lifetime $\tau_{D}$ estimated from the Drude peaks.  In Fig.~\ref{fig4}(c) we compare different hot-carrier scattering times for $\mu=-100$ meV. Upon increasing $\eta$ the sharp features in Fig.~\ref{fig4}(b) are broadened, turning the sharp peak into a shoulder, which was observed in the experimental data. When  $\eta = 100-150$ meV the resulting optical conductivity falls close to our experimental data, including the upturn at $0.4$ eV, associated to the activation of the broadened transitions around the $R$ point, which agrees with the experimental data. Note that the large disorder scale is similar to the spin-orbit coupling ($\approx$ 100 meV) and, therefore,  washes out any feature narrower than 100 meV and justifies our discussion based on a tight-binding model without spin-orbit coupling.

We note that despite the general agreement below $0.5$ eV, the intuitive tight-binding calculations deviate from the data above $0.5$ eV. This is likely due to the tight-binding model's known limitations, which fails to accurately capture the band structure curvature and orbital character at other high-symmetry points such as the $M$ point, which is a saddle point. These limitations will also play a role in our discussion of the CPGE.

To refine our understanding of these aspects and to further examine the role of spin-orbit coupling, we have used the DFT method to calculate the optical conductivity, including spin-orbit coupling on a wider frequency range up to $\hbar\omega=4$ eV (See Methods). The optical conductivity we obtained is compared to our data in Fig.~\ref{fig5}(a) for a wide range of frequencies, and in Fig.\ref{fig5}(b) within a low energy frequency window. The smaller broadening factors, $\eta=10,50$ meV, reveal fine features due to spin-orbit coupling such as the peaks at low energies, due to the spin-orbit splitting of the $\Gamma$ and $R$ points (see Fig. \ref{fig3}(b)). These features are absent in the data as they are smoothened as the broadening is increased (see Fig.~\ref{fig5}(a)), consistent with the tight-binding model discussion above. As shown in Fig.~\ref{fig5}(b), the low-energy conductivity below 0.2 eV is better explained if the chemical potential is at - 30 meV. The dip at around 0.5 eV, seen in Fig.~\ref{fig5}(b) for low broadening, is filled with spectral weight as broadening is increased, consistent with our tight-binding calculations. Note that the DFT calculation underestimates the conductivity in the range of 0.2 - 0.4 eV, which is probably because the contribution of surface arcs is not considered \cite{changPRL2020}.

At higher energies the DFT calculation recovers a peak at $\hbar\omega \approx 1$ eV, seen also in our data (see Fig.~\ref{fig5}(a)). If the broadening is small ($\eta$=10 meV), the peak is sharper compared to the measurement and exhibits a double feature with a shoulder around 0.85 eV and a sharp peak around 1.08 eV . Although smoothened by disordered, this double feature is consistent with the need of two Lorentzian functions (Lorentz 1 and Lorentz 2 in Fig.~\ref{fig2}(a)) to model this peak phenomenologically with Eq.~\eqref{DrudeLorentz}. Fig.~\ref{fig3}(b) shows that the excitation energy at the $M$ point is around 0.7 eV. In a recent work on CoSi, in the same space group 198 as RhSi, a momentum-resolved study shows that the dominant contribution of this Lorentzian function (Lorentz 1) arises from the saddle point at $M$~\cite{xuarXiv2020}. The peaks around 1.1 eV and 2.5 eV could arise from other saddle points with a gap size larger than that at the $M$ point. 

Overall, the curves with broadening factor $\eta=100$ meV and with chemical potential below the nodes at $\Gamma$ in both the tight-binding and DFT calculations show a good qualitative agreement with the experimentally measured curve in a wide frequency range. This observation determines approximately the hot-carrier lifetime in RhSi to be $\tau=\hbar/\eta\approx 6.6$ fs. \\

\textbf{CPGE measurement} 

\begin{figure*}
\includegraphics[width=0.95\textwidth]{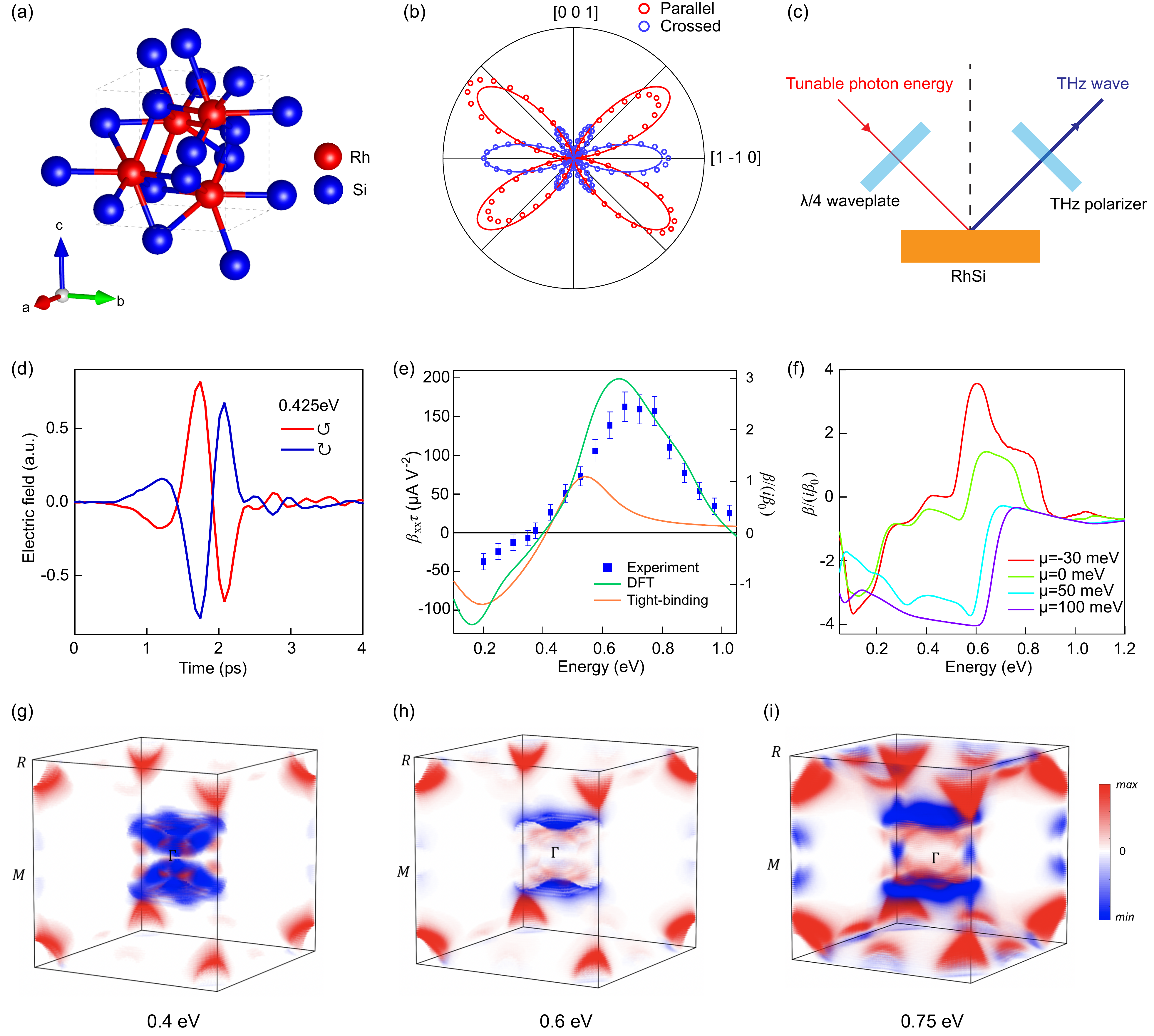}
\caption{\label{fig6} 
\textbf{CPGE measurement and calculation.} {(a) Crystal structure of RhSi. (b) Second harmonic generation  measured on the (110) surface of RhSi crystal at near-normal incidence. Solid red and blue lines are the best fit of the experiment data. Open circles are data measured in the parallel- and crossed- polarizer geometry. 
(c) A schematic representation of the experimental geometry for the THz emission spectroscopy. (d) Detected emitted THz pulses in the incident plane from RhSi at the 45-degree incidence of left-handed and right-handed circularly polarized light at the photon energy of 0.425 eV.
(e) Experimental CPGE spectrum of RhSi (blue squares) compared to first-principles at $\mu=-30$ meV and $T=300$ K and tight-binding calculations. Both calculations use a broadening $\eta=100$ meV.
(f) First-principles calculation of the trace of the CPGE components in units of the quantization constant $\beta_0 = \pi e^3 /h^{2}$ with spin orbit coupling at different chemical potential $\mu$ from  from the zero energy defined in Fig.~\ref{fig3}(a)(b) at $T=0$ K, and a broadening parameter $\eta=10$ meV. While $\mu=0$ and -30 meV sit below the $\Gamma$ threefold node (see Fig.~\ref{fig3}(a) and (b)), the two remaining chemical potentials lie above the $\Gamma$ node. (g-i) Momentum resolved contribution of the CPGE photocurrent at 0.4 eV, 0.6 eV and 0.75 eV for the green curve in (e). The contribution is shown by the red-blue color scale with red being positive and blue being negative.}}
\end{figure*}

As a first step of the CPGE experiment, we determined the high symmetry axes, [0,0,1] and [1,-1,0] directions of the RhSi (110) sample (See Fig.~\ref{fig6}(a)) respectively by second harmonic generation (SHG). To stimulate SHG, pulses of 800 nm wavelength were focused at near-normal incidence to a spot with a 10- $\mu$m diameter on the  (110) facet \cite{wuNatPhys2017}. Fig.~\ref{fig6}(b) shows the polar patterns of SHG as a a function of the polarization of the linear light in the co-rotating parallel-polarizer (red) and crossed-polarizer (blue) configurations\cite{wuNatPhys2017,patankarPRB2018}.  The solid lines are the fit constrained by the point group symmetry with only one non-zero term $\chi^{(2)}_{xyz}$ and the angle dependence of the SHG are:
 
 \begin{equation}
I_{parallel}(\theta)=36|\chi^{(2)}_{xyz}|^2\cos^4\theta\sin^2\theta,
\end{equation}

\begin{equation}
I_{cross}(\theta_1)=4|\chi^{(2)}_{xyz}|^2(2\cos\theta\sin^2\theta-\cos^3\theta)^2.
\end{equation}
 
\noindent where $\theta$ is the angle between the polarization of the incident light and the [1,-1,0] axis.

 Next, we perform THz emission spectroscopy to measure the longitudinal CPGE in RhSi (See Methods).  As shown in Fig.~\ref{fig6}(c), an ultrafast circularly-polarized laser pulse is incident on the sample at 45 degrees to generate a transient photocurrent. Due to the longitudinal direction of the CPGE, the transient current flows along the light propogation direction inside RhSi, and therefore, in the incident plane\cite{dejuanNatComm2017,changPRL2017,flickerPRB2018}. The THz electric fields radiated by the time-dependent photocurrent is collected and measured by a standard electro-optical sampling method with a ZnTe detector\cite{shan2004terahertz}.  The component in the incident plane is measured by placing a THz wire-grid polarizer before the detector.  Fig.~\ref{fig6}(d) shows a typical response of the component of emitted THz pulses in the incident plane under left-handed and right-handed circularly polarized light at an incident energy of 0.425 eV. The nearly opposite curves demonstrate the dominating CPGE contribution to the photocurrent with an almost vanishing linear photogalvanic effect (LPGE) at this incident energy. The  CPGE contribution can be extracted by taking the difference between the two emitted THz pulses under circularly polarized light with opposite helicity. During our measurement, the [0 0 1] axis  is kept horizontally in the lab, even though the CPGE signal does not depend on the crystal orientation due to the cubic crystalline structure.

  In order to measure the amplitude of the CPGE  photocurrent, we use a motorized delay stage to move a standard candle ZnTe at the same position  to perform the THz emission experiment right after measuring RhSi for every  photon energy between 0.2 eV to 1.1 eV \cite{SotomePNAS19}. ZnTe is a good benchmark due to its relatively flat frequency dependence on the electric-optical sampling coefficient for photon energy below the gap~\cite{Cabrera1985, boyd2003nonlinear}. The use of ZnTe circumvents assumptions regarding the incident pulse length, the wavelength dependent focus spot size on the sample, and the calculation of collection efficiency of the off-axis parabolic mirrors \cite{SotomePNAS19, niCoSi2020}. The details of the derivation can be found in Ref. \onlinecite{niCoSi2020} and this method was previously used in the shift current measurement on a ferroelectric insulator \cite{SotomePNAS19}. A spectrum of CPGE photo-conductivity as a function of incident photon energy is shown in Fig.~\ref{fig6}(e) in units of $\rm{\mu A/V^2}$ (squares). Upon decreasing the incident photon energy from 1.1 eV to 0.7 eV, we observe a rapid increase of CPGE response with a peak value of  163 ($\pm$19) $\rm{\mu A/V^2}$ at 0.7 eV. The features of this line resemble those observed in Ref.~\onlinecite{reesarxiv2019}. Further decreasing the photon energy from 0.7 eV to 0.2 eV, the CPGE conductivity displays a sharp drop with a striking sign change at 0.4 eV, which was not seen before as the lowest photon energy measured in a previous study was around 0.5 eV \cite{reesarxiv2019}. Interestingly, the peak photoconductivity at 0.7 eV is  much larger than the photo-galvanic effect in BaTiO$_3$\cite{feiPRB2020}, single-layer monochalcogenides \cite{rangelPRL2017,feiPRB2020} other chiral crystals\cite{zhangPRB2019} and it is comparable to  the colossal bulk-photovoltaic response in TaAs\cite{osterhoudtNatMat2019}. It is also one order of magnitude larger than the previous study on RhSi \cite{reesarxiv2019} probably due to a larger hot-carrier lifetime. Interestingly, the sign change at 0.4 eV  was not predicted in previous theory studies either\cite{changPRL2017,flickerPRB2018,deJuan:2020jm}. The quantized CPGE below 0.7 eV predicted in RhSi in previous theory studies \cite{changPRL2017,flickerPRB2018,deJuan:2020jm} is absent in the experiment. Note that CPGE from the surface Fermi arcs, which might exist below 0.5 eV on the (110) facet, is generally one order of magnitude smaller than the bulk contribution. It is better to be detected under normal incidence and therefore not the focus of this work  \cite{changPRL2020}.\\

\textbf{CPGE calculation} 

In order to understand the absence of quantized CPGE and the origin of the sign change in our CPGE data, we have calculated the CPGE response, $\beta_{ij}$,  using a first-principle calculation via FPLO (full-potential local-orbital
minimum-basis) as DFT captures the curvature of the flat bands at $\Gamma$ and  the saddle point $M$~\cite{Koepernik1999, perdew1996} (See Methods). Due to cubic symmetry, the only finite CPGE component is $\beta_{xx}$\cite{flickerPRB2018}. In our convention, the tensor $\beta_{ij}$ determines the photocurrent rate. When the hot-carrier lifetime $\tau$ is short compared to the pulse width,  the total photocurrent is given $\beta_{xx}\tau$\cite{deJuan:2020jm}. $\beta_{xx}$ is directly calculated from the band structure at $\mu=-30$ meV and we assume a constant $\tau$ as a function of energy. $\tau$ is the only fitting parameter to match both the peak and width in the CPGE current.

In Fig.~\ref{fig6}(e) we plot $\beta_{xx}\tau$, with the hot-carrier scattering time corresponding to a broadening $\eta= \hbar/ \tau=100$ meV ($\tau\approx 6.6$ fs) calculated using the DFT ($\mu=-30$ meV, $T=300$ K). The DFT calculation captures quantitatively the features seen in the CPGE data: the existence of a peak around $0.7$ eV, its width, and the sign change of the response. Together they support the conclusion that the chemical potential lies below the $\Gamma$ node (see Fig.~\ref{fig3}), consistent with the features of the optical conductivity. Fig. \ref{fig6} (g-i) shows the momentum-resolved contribution to the CPGE current at different incident photon energies. Below 0.6 eV, the main contributions are centered around the $R$ and $\Gamma$ points with opposite signs while the $M$ point is turned on at 0.75 eV. The sign change at 0.4 eV is due to the turn on of the excitations at $R$ with an opposite sign in $\beta_{xx}$, which was also derived in a simpler $k \cdot p$ model recently \cite{niCoSi2020}. Note that at 0.4 eV, the $R$ point already contributes to the CPGE due to the large broadening $\sim$ 100 meV in this material. The certain remaining differences between the data and the calculations suggests that a constant, energy independent hot-carrier scattering time $\tau$ might be an oversimplified phenomenological model for the disorder. In general, the hot-carrier scattering time is energy and momentum dependent~\cite{Konig2017} and  including these effects might give even better agreement.

It is illustrative to compare these results, especially the striking sign change, with a four-band tight-binding calculations for the CPGE as the latter is the simplest model to capture both the multifold fermions at the $\Gamma$ and $R$ points. Following Ref.~\onlinecite{flickerPRB2018} we computed the CPGE with parameters $\eta=100$ meV and $\mu=-100$ meV (Fig.~\ref{fig6}(e) orange line), which match the optical conductivity (see Fig.~\ref{fig4}(c), solid line), but it underestimates the position of the peak and the overall magnitude of the CPGE. This is mainly attributed to the failure of the tight-binding to capture the $M$ point. However, it shows the overall peak-dip structure of the response and its sign change around 0.4 eV, which is contributed from the negative chemical potential at $\Gamma$. By lowering the chemical potential, the tight-binding result can be made to match  the data, paying the price that the optical conductivity will no longer be reproduced.

We end by discussing the possibility of observing a quantized CPGE in this sample. In Fig.~\ref{fig6}(f) we show the effect of changing the chemical potential on the DFT  calculated CPGE tensor trace $\beta=\mathrm{Tr}[\beta_{ij}]=3\beta_{xx}$ in units of the quantization constant $\beta_0 = \pi e^3 /h^{2}$. For an ideal multifold fermion with linear dispersion, and taking into account spin-degeneracy, the CPGE is expected to be quantized to a Chern number of four, as $\beta = 4i\beta_0$, corresponding to the total charge of the nodes at $\Gamma$\cite{flickerPRB2018}, $C=4$. A finer analysis and density functional theory calculations\cite{flickerPRB2018,deJuan:2020jm} indicates that, unlike in the case of Weyl nodes, quadratic corrections can spoil quantization beyond the linear dispersion regime in multifold fermion materials. Upon decreasing the broadening to 10 meV and changing the Fermi level by  100 meV compared to the chemical potential found in our DFT calculations,  a narrow frequency window around  0.6 eV emerges with a close-to-quantized value, shown as purple line in Fig.~\ref{fig6}(f). We note that when the chemical potential is above the nodes, there is no sign change below 0.7 eV.\\

In conclusion, we have established a consistent picture of the optical transitions in RhSi using a broad set of theoretical models applied to interpret the linear and nonlinear optical responses. Our data is explained if the chemical potential crosses a large hole-like band at $\Gamma$, and with a relatively short hot-carrier lifetime $\approx 4.4-6.6$ fs. The combined analysis of both linear and nonlinear responses illustrates the crucial role played by the curvature of the flat-band at the $\Gamma$ point and the saddle point at $M$.

Interband optical conductivity shows two quasi-linear regions where the conductivity increases smoothly with frequency and a slope change around 0.4 eV. The slope in the first region is determined by a disorder-broadened  contribution associated with a threefold fermion at the $\Gamma$ point. The slope in the second region is determined by the onset of a broadened $R$ point conductivity. 

The circular photogalvanic effect exhibits a sign change close to $\hbar\omega\sim 0.4$ eV, and  a non-quantized peak at $\approx 0.7$ eV. The magnitude of the CPGE response is approximately captured by our density functional theory (DFT) calculations for a wide range of frequencies.  Lastly, our calculations suggest that by electron-doping RhSi by $\approx 100$ meV, a close-to-quantized value could be observed in a narrow energy window around 0.6 eV, if the hot-carrier scattering time is significantly increased. To realize the quantized CPGE, it would be also desirable to identify a material candidate with smaller spin-orbit coupling than RhSi \cite{le2020ab,niCoSi2020}.

Our systematic methodology can be applied to other non-centrosymmetric topological materials \cite{bradlynScience2016,changNatMat2018} to reveal signature of topological excitations. We observed THz emssion in the mid-infared regime (0.2 eV - 0.5 eV). We expect that the development of broadband THz emission spectroscopy provides the opportunity to reveal bulk photovoltanic \cite{SotomePNAS19, siricaPRL2019, gaoNatComm2020} and spintronic responses \cite{nvemecnatphys18} in a low-energy regime and also the possibility of probing Berry curvature in surface-state photo-galvanic effect in topological insulators \cite{hosurPRB2011}. \\

\textbf{Methods} 

\textbf{Crystal growth} The high-quality single crystal of RhSi was grown by the Bridgeman method~\cite{sanchez2019topological}. 2 mm$\times$5 mm large RhSi with a (110)  facet is used in this study.

\textbf{Optical conductivity measurement} The in-plane reflectivity $R(\omega)$ was measured at a near-normal angle of incidence using a Bruker VERTEX 70v FTIR spectrometer with an \emph{in situ} gold overfilling technique~\cite{Homes1993}. Data from 30 to 12\,000\icm\ ($\simeq$ 4~meV to 1.5~eV) were collected at different temperatures from 10 to 300~K with a ARS-Helitran cryostat. The optical response function in the near-infrared to the ultraviolet range (4\,000 -- 50\,000\icm) was extended by a commercial ellipsometer (Woollam VASE) in order to obtain more accurate results for the Kramers-Kronig analysis of $R(\omega)$~\cite{Dressel2002}. The beam is focused down to 2 mm and not polarized as the conductivity is isotropic due to cubic symmetry.

\textbf{THz emission experiment} The THz emission experiment is performed at dry air environment with relative humidity less than 3$\%$ at room temperature. An ultrafast laser pulse is incident on the sample at 45 degrees to the surface normal and is focused down to  1 mm$^2$ to induce  THz emission. The THz pulse is  focused by a pair of 3-inch off-axis parabolic mirrors on the ZnTe (110) detector. The temporal THz electric field can be directly measured with a gated probe pulse of 1.55 eV and 35 fs duration\cite{shan2004terahertz}.  The polarization of incident light is controlled by either a near-infrared achromatic or a mid-infrared quarter-wave plate. A wire-grid THz polarizer is utilized to pick out the THz electric field component in the incident plane. The photon energy of the incident light is tunable from 0.2 eV to 1.1 eV by an optical parametric amplifier and difference frequency generation. Pulse energy of \SI{12}{\micro\joule} is used for 0.4-1.1 eV and \SI{6}{\micro\joule} is used for 0.2-0.4 eV. The repetition rate of the laser used is 1 kHz.

\textbf{Four-band tight-binding model}
In the main text we use a four-band tight-binding model introduced in Ref.~\onlinecite{changPRL2017} and further expanded in Ref.~\onlinecite{flickerPRB2018}. Since this model has been extensively studied before, we briefly review its main features relevant to this work. Our notation and the model is detailed in appendix F of Ref.~\onlinecite{SanchezMartinez:2019he}.

Without spin-orbit coupling,  the four-band tight-binding model is determined by three-material dependent parameters, $v_1$, $v_p$, and $v_2$. By fitting the band structure in Fig~\ref{fig3}(b) we set $v_1 = 1.95$, $v_p = 0.77$ and $v_2 = 0.4$. Our DFT fits deliver parameters that differ from those obtained in Ref.~\onlinecite{changPRL2017}, subsequently used in Ref.~\onlinecite{SanchezMartinez:2019he}. Our current parameters result in a better agreement to the observed optical conductivity and circular photogalvanic effect data. Additionally, we rigidly shift the zero of energies of the tight-binding model by $0.78$ meV with respect to Ref.~\onlinecite{SanchezMartinez:2019he} to facilitate comparison with the DFT calculation. As described in Ref.~\onlinecite{flickerPRB2018} we additionally incorporate the orbital embedding in this model, which takes into account the position of the atoms in the unit cell. It amounts to a unitary transformation of the Hamiltonian which depends on the atomic coordinates through a material dependent parameter $x$, where $x_{\mathrm{RhSi}}=0.3959$ for RhSi~\cite{flickerPRB2018,SanchezMartinez:2019he}. 

\textbf{Details of the optical conductivity calculations}

Since RhSi crystallizes in a cubic lattice it is sufficient to calculate one component of the real part of the longitudinal optical conductivity, $\sigma_{1}(\omega)$, given by the expression~\cite{Sipe93} 
\begin{align}
\label{eqn:optcond}
\sigma_{1}(\omega)=-\frac{\pi e^2\hbar}{V}\sum_{\mathbf{k},m\neq n}\frac{\hbar\omega f_{nm}}{\epsilon_{nm}^2}|v^x_{nm}|^2 \mathcal{L}_\tau(\epsilon_{nm}-\hbar\omega),
\end{align}
defined for a system of volume $V$ described by a Hamiltonian $H$ with eigenvalues and eigenvectors $\epsilon_n$ and $\ket{n}$, respectively, and a velocity matrix element $v^x_{nm} = \frac{1}{\hbar}\bra{n} \partial_{k_{i}}H\ket{m}$. The chemical potential $\mu$ and temperature $T$ enter through the difference in Fermi functions $f_{nm}=f_n-f_m$, and we define $\epsilon_{nm}=\epsilon_{n}-\epsilon_{m}$. We have replaced the sharp Dirac delta function that governs the allowed transitions by a Lorentzian distribution $\mathcal{L}_\tau(\omega)=\frac{1}{\pi} \frac{\eta}{\omega^2+\eta^2}$ to phenomenologically incorporate disorder with a constant hot-carrier scattering time $\tau= \hbar/\eta$.

The DFT calculations of the optical conductivity are also performed via Eq.~\eqref{eqn:optcond}, with the ab-\textit{initio} tight binding Hamiltonian constructed from DFT calculations. We use a dense $300\times300\times 300$ momentum grid for the small broadening factor $\eta=10$ meV, and a $200\times200\times 200$ momentum grid for  broadenings $\eta \geq 50$ meV.

\textbf{Details of the CPGE calculations}
For our \textit{ab-initio} CPGE calculations we projected the
\textit{ab-initio} DFT Bloch wave function into atomic-orbital-like Wannier functions~\cite{first_wannier}.
To ensure the accuracy of the Wannier projection, we have
included the outermost $d$, $s$, and $p$ orbitals for transition
metals ($4d$, $5s$, and $5p$ orbitals for Rh) and the
outermost $s$ and $p$ orbitals for main-group elements. Based on
the highly symmetric Wannier functions, we constructed an
effective tight-binding model Hamiltonian and calculated the
CPGE evaluating~\cite{Sipe93,deJuan:2020jm}
\begin{align}
\nonumber
{\beta_{ab}}(\omega) &= {\frac{i\pi e^3}{4\hbar}}  \int_{\rm BZ} \frac{{d\mathbf{k}}}{(2\pi)^3} {\sum_{n>m}} f_{nm}\epsilon^{bcd} \\
&\times \Delta^a_{mn}{\rm Im} [r^d_{nm}r^c_{mn}] \mathcal{L}_\tau(\epsilon_{nm}-\hbar\omega), \label{eq:injection}
\end{align}
using a dense $480\times480\times 480$ momentum grid. Here  $\Delta^a_{mn}\equiv\partial_{k_a}\epsilon_{mn}/\hbar$, and $r_{mn}^a \equiv i \langle m|\partial_{k_a}n\rangle = v_{nm}/i\epsilon_{nm}$ is the interband transition matrix element or off-diagonal Berry connection. As for the optical conductivity, the chemical potential and temperature are considered via the Fermi-Dirac distribution $f_n$, and the Lorentzian function $\mathcal{L}_\tau(\omega)$ accounts for a finite hot-carrier scattering time.

\textbf{Data and materials availability:} All data needed to evaluate the conclusions are present in the paper. Additional data related to this paper could be requested from the authors.

\textbf{Acknowledgements}
We thank  G. Chang and Z. Fang for helpful discussions. Z.N. and L.W. are supported by Army Research Office under Grant W911NF1910342.  J.W.F.V. is supported by a seed grant from NSF MRSEC at Penn under the Grant DMR-1720530. B.X. and C.B. were supported by the Schweizerische Nationalfonds (SNF) by Grant No. 200020-172611.  M. A. S. M acknowledges support from the European Union's Horizon 2020 research and innovation programme under the Marie-Sklodowska-Curie grant agreement No. 754303 and the GreQuE Cofund programme. A. G. G. is supported by the ANR under the grant ANR-18-CE30-0001-01 (TOPODRIVE) and the European Union Horizon 2020 research and innovation programme under grant agreement No. 829044 (SCHINES). F. J. acknowledges funding from the Spanish MCI/AEI through grant No. PGC2018-101988-B-C21. Y.Z. is currently supported by the the DOE Office of Basic Energy Sciences under Award desc0018945 to Liang Fu.  Y.Z., K.M. and C.F. acknowledge financial support from the European Research Council (ERC) Advanced Grant No.742068 ``TOP-MAT'' and Deutsche Forschungsgemeinschaft (Project-ID 258499086 and FE 633\/30-1). This research was supported in part by the National Science Foundation under Grant No. NSF PHY11-25915. The DFT calculations were carried out on the Draco cluster of MPCDF, Max Planck society.\\

\textbf{Author Contribution:} L.W. conceived the project and coordinated the experiments and theory.  Z.N. performed the THz emission experiments and analyzed the data with L.W..   B.X. and C.B. performed the optical conductivity measurement, and analyzed the data together with L.W. J.W.F.V. fitted the DFT band structure. M.A.S.M and A.G.G. performed the tight-binding calculation. Y.Z. performed the DFT calculation. K.M and C.F. grew the crystals. L.W., A.G.G. and F.J. interpreted the data jointly with the calculations.  L.W. and A.G.G. wrote the manuscript with inputs from B.X., Y.Z. and N.Z..  All authors edited the manuscript.

Z.N. and B.X. contributed equally to this project.

\textbf{Competing interests:} The authors declare that they have no competing interests.

\bibliographystyle{naturemag}

\end{document}